\documentclass[preprint,11pt,nonatbib]{elsarticle}

\usepackage[T1]{fontenc}
\usepackage[utf8]{inputenc}
\usepackage{lmodern}
\usepackage[margin=1in]{geometry}
\usepackage{booktabs}
\usepackage{enumitem}
\usepackage{graphicx}
\usepackage[hidelinks]{hyperref}
\usepackage[backend=biber, style=numeric]{biblatex}
\journal{Journal of Pathology Informatics}

\begin{document}

\begin{frontmatter}

\title{STEP: A Modular Silent Trial Engine for Operational Evaluation of Digital Pathology AI in Routine Workflow}

\author[inst1,inst2]{Gabriele Campanella}
\cortext[cor1]{Corresponding author}
\ead{gabriele.campanella@mssm.edu}

\author[inst3]{Matthew Croken}
\author[inst3]{Olga Lukatskaya}
\author[inst3]{Jane Houldsworth}
\author[inst3]{Ricky Kwan}
\author[inst4]{Peter Sch\"uffler}
\author[inst5]{Chad Vanderbilt}

\address[inst1]{Windreich Department of AI and Human Health, Icahn School of Medicine at Mount Sinai, New York}
\address[inst2]{Hasso Platner Institute at Mount Sinai, Icahn School of Medicine at Mount Sinai, New York}
\address[inst3]{Department of Pathology, Icahn School of Medicine at Mount Sinai, New York}
\address[inst4]{Institute of Pathology, Technical University of Munich, Munich}
\address[inst5]{Department of Pathology, Memorial Sloan Kettering Cancer Center, New York}

\begin{abstract}
\textbf{Background:} Prospective silent trials provide an important bridge between retrospective validation of artificial intelligence (AI) models and their use in clinical care by evaluating model performance and operational reliability on live clinical data without influencing patient management. In computational pathology, however, conducting silent trials requires integration across laboratory information systems, digital pathology infrastructure, computational resources, and model inference pipelines, and these workflows are often implemented using application-specific software.

\textbf{Objective:} We developed the Silent Trial Engine for Pathology (STEP), a reusable software platform for orchestrating prospective silent trials of computational pathology AI models across heterogeneous clinical and computational environments.

\textbf{Methods:} STEP separates common silent-trial orchestration from institution-specific data access and compute infrastructure through modular adapter interfaces. The platform supports scheduled case discovery, per-slide inference submission, deterministic idempotency, failure recovery, result and ancillary-data ingestion, persistent trial and run state, and audit logging. Compute adapters support local execution and high-performance computing environments using LSF and Slurm. STEP was deployed at three institutions to support prospective silent evaluation of EAGLE, an AI model for predicting \textit{EGFR} mutation status from hematoxylin and eosin-stained whole-slide images.


\textbf{Conclusions:} STEP provides a reusable orchestration layer for prospective silent evaluation of computational pathology AI models. By separating trial-level workflow logic from site-specific data and compute integrations, STEP enables the same core execution framework to operate across heterogeneous pathology environments while maintaining durable and auditable trial state. This approach may reduce duplicated engineering effort and facilitate systematic real-world evaluation of computational pathology AI before interventional clinical deployment.
\end{abstract}


\end{frontmatter}

\section{Introduction}

Advances in artificial intelligence (AI) and computer vision have fueled rapid growth in computational pathology, with applications ranging from detection and segmentation to workflow automation and the development of prognostic and predictive biomarkers\cite{Marra2025PathologyAI,Aggarwal2025RealityCheck,Bera2019DigitalPathology}. Publications in the field have increased threefold over the past five years, reflecting its rapid scientific maturation\cite{Aggarwal2025RealityCheck}. The convergence of digital pathology infrastructure, increasingly capable AI models, and mounting clinical demand has created an opportunity to move computational pathology beyond algorithm development toward meaningful clinical implementation.

Despite this technical progress, clinical adoption of AI in pathology remains limited. Compared with radiology and cardiology, where dozens of FDA-approved AI tools are in routine use, pathology accounts for less than 0.5\% of FDA-approved software-as-a-medical-device products\cite{Aggarwal2025RealityCheck,Benjamens2020FDADevices}. Even among the few FDA-cleared pathology applications, including Paige Prostate\cite{FDA2021PaigeProstate}, PathAI/Roche Digital Pathology Dx\cite{Roche2024DigitalPathologyDx}, Concentriq AP-Dx\cite{Proscia2024ConcentriqAPDx}, and ArteraAI Prostate\cite{Artera2025DeNovo}, clinical implementation has remained limited in part because of incomplete real-world validation and an evolving regulatory landscape\cite{Aggarwal2025RealityCheck,ReisFilho2023ImplementationAI,Lam2022RCTAI}. Existing approved applications also address a relatively small number of narrowly defined diagnostic use cases, leaving most routine pathology workflows without clinically validated AI solutions. These limitations underscore that technical performance alone is insufficient to translate computational pathology into widespread clinical impact.

A central barrier to translation is the lack of prospective evidence demonstrating that AI systems function reliably and provide value within real-world clinical workflows. Although hundreds of computational pathology studies have reported promising diagnostic, prognostic, and predictive models, most evaluations remain retrospective and focus on algorithm development or external validation. Retrospective studies can characterize model discrimination and generalizability across existing datasets, but they cannot fully establish whether an end-to-end AI system will function as intended when deployed prospectively. Models may encounter shifts in patient populations and data distributions, discrepancies between research and production data pipelines, or dependencies on variables that appear available retrospectively but are unavailable at the time of inference\cite{Corbin2023DEPLOYR}. Moreover, clinical readiness depends not only on predictive accuracy but also on whether eligible patients can be identified in real time, required data can be accessed, inference can be executed reliably, and predictions can be returned within clinically actionable time windows.

Prospective ``silent trials'' have therefore emerged as an important bridge between retrospective model validation and interventional clinical evaluation\cite{Kwong2022SilentTrial}. In a silent trial, a frozen AI system executes prospectively on live clinical data within its intended deployment environment, while its predictions remain invisible to clinicians and do not influence patient care. This approach enables evaluation of model performance, operational reliability, data availability, and temporal actionability under authentic clinical conditions without exposing patients to AI-guided decisions. Silent trials can consequently identify failure modes that are difficult or impossible to detect retrospectively and generate real-world evidence before progression to interventional studies.

The use of this paradigm is increasing across clinical AI. A recent scoping review identified 75 silent evaluations of medical AI across 16 countries, demonstrating growing adoption of prospective silent evaluation as a translational methodology\cite{Tikhomirov2026SilentTrials}. Its application to computational pathology, however, remains exceptionally limited. To our knowledge, the prospective deployment of EAGLE\cite{Campanella2025EAGLE} from our team was the only silent trial of a hematoxylin and eosin (H\&E)-based computational pathology model identified in this review. This gap is particularly consequential for pathology, where deployment requires coordination across laboratory information systems, digital pathology infrastructure, whole-slide image storage, computational resources, and clinical workflows\cite{Schueffler2021}.

Conducting prospective silent trials within this environment introduces substantial technical and operational challenges. Implementations may require continuous identification of eligible cases, retrieval and validation of clinical and imaging data, orchestration of model inference, management of computational resources, tracking of failures and retries, and aggregation of predictions for subsequent analysis. These functions are frequently implemented through application-specific scripts and pipelines, limiting maintainability, reproducibility, and auditability. Although institutional data architectures and computational environments vary, the underlying logic required to execute a silent trial is largely shared: cases must progress through a defined sequence of states from identification and data acquisition through inference and completion, while preserving sufficient provenance to reconstruct the execution history of each case.

This common structure motivates a reusable orchestration layer for prospective computational pathology studies. In this technical note, we present STEP (Silent Trial Engine for Pathology), a modular platform designed to facilitate the execution of pathology AI silent trials across heterogeneous clinical and computational environments. STEP provides structured support for case orchestration, scheduling, compute portability, failure recovery, and auditable state transitions while separating trial execution logic from institution-specific data access and model implementation. By abstracting these recurring components, STEP aims to reduce the engineering burden of prospective deployment and provide a reproducible framework for generating real-world evidence during the translation of computational pathology AI from retrospective validation toward clinical evaluation.

\section{STEP System Design}

STEP was designed as a reusable orchestration layer for prospective, non-interventional evaluation of pathology AI models. Its design reflects the operational requirements that recur across silent-trial deployments: scheduled and on-demand discovery of eligible cases, slide-level inference submission, ingestion of delayed ancillary information, durable persistence of run and result state, audit logging of operational events, and concurrent support for multiple trials with independent schedules and configuration. Rather than encoding these concerns in trial-specific scripts, STEP separates the common control flow of a silent trial from the local details of data access, compute infrastructure, and result production.

\begin{figure}[h!]
    \centering
    \includegraphics[width=\textwidth]{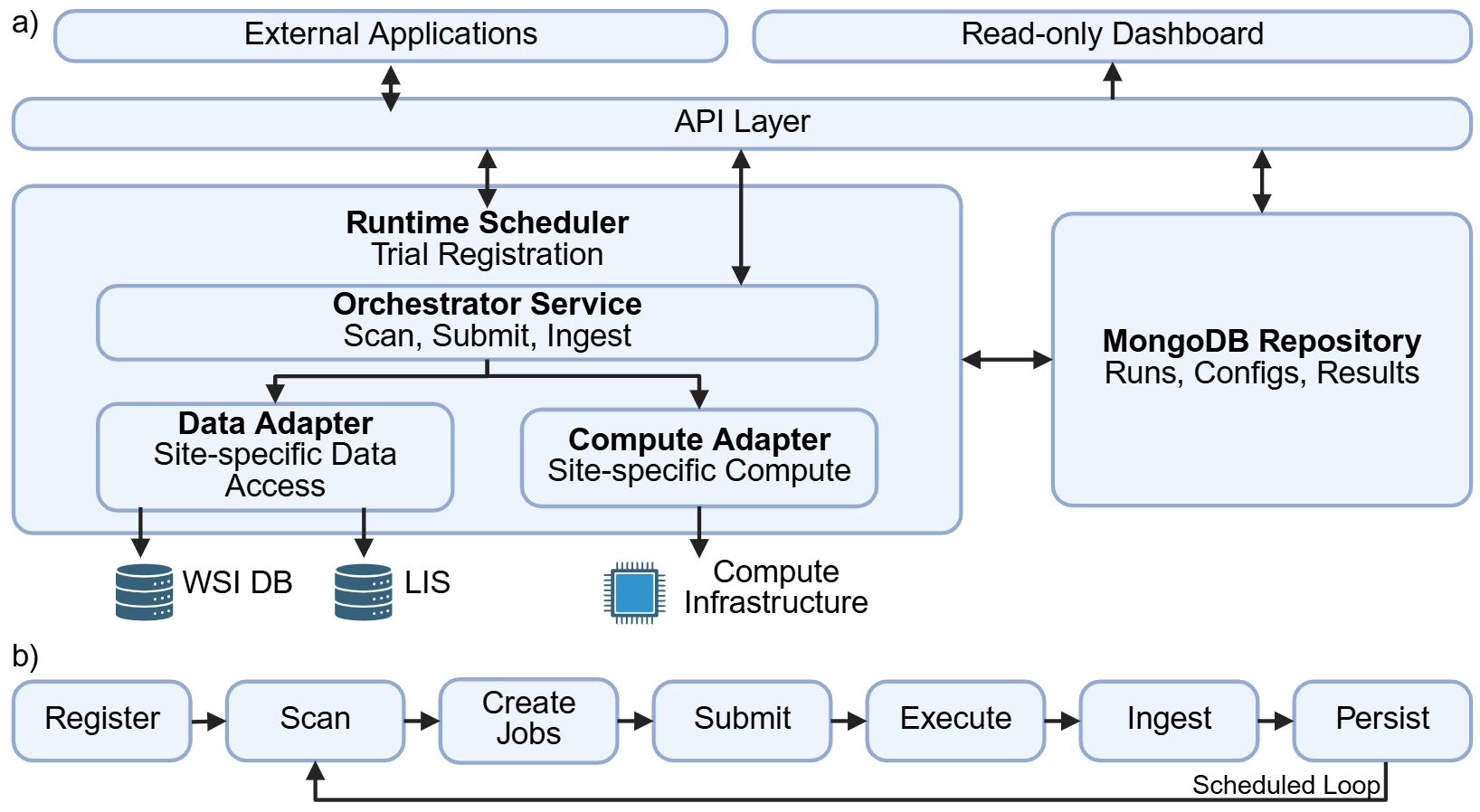}
    \caption{\textbf{STEP architecture and silent-trial workflow.}
(a) STEP separates common trial orchestration from site-specific data access and computational infrastructure. The runtime scheduler manages trial registration and scheduled execution, while the orchestrator service coordinates case scanning, inference submission, and result ingestion. Data and compute adapters provide interfaces to local whole-slide image (WSI) repositories, laboratory information systems (LIS), and computational resources. Trial configuration, run state, and results are maintained in the MongoDB repository and exposed through the API layer and read-only dashboard.
(b) Operational life cycle of a STEP trial. Following registration, STEP periodically scans for eligible cases, creates and submits inference jobs, executes model inference through the configured compute adapter, ingests completed results, and persists trial state. The scan-to-persistence sequence is repeated according to the configured trial schedule.}
    \label{fig:pipeline}
\end{figure}

The package is organized as a set of layered components (Figure~\ref{fig:pipeline}a). The FastAPI application provides the external control plane; the runtime layer materializes trial-specific schedulers and orchestrators; the orchestration service implements the scan, submit, and ingest loop; adapters connect the common workflow to local data sources, compute backends, and ancillary-data feeds; and a MongoDB repository stores trial configuration, case records, jobs, results, and audit events. In routine operation, a trial is registered through the API, scheduled jobs are created for case scanning and result ingestion, candidate cases and slides are retrieved from the configured data source, one inference run is submitted per eligible slide, and completed result files are normalized back into the persistent run record.

\subsection{Architecture}

The runtime layer (\texttt{step/runtime.py}) is responsible for assembling the application context. It binds the MongoDB repository, adapter factories, and scheduler service, and it maintains an in-memory registry of active trials. Registration produces a trial-scoped orchestrator and a set of recurring scheduler jobs for case scanning, result ingestion, and ancillary-data ingestion. Because registered trial configurations are also stored in MongoDB, the service can restore active trials at startup and resume scheduled operation after a process restart.

The orchestration layer (\texttt{step/services/orchestrator.py}) contains the central silent-trial loop. During a scan cycle, it queries the configured data source over the active lookback window, upserts discovered cases, evaluates each slide for prior submission, writes an inference payload, and delegates execution to the selected compute adapter. During ingestion, it polls the compute adapter for completed result artifacts, saves normalized result objects, updates job status, and records audit events. The same orchestrator also supports failed-run retry and optional payload cleanup after terminal runs, which keeps the retry and retention behavior close to the run state it modifies.

Persistence is implemented through a MongoDB repository (\texttt{step/repositories/mongo\_repository.py}). STEP stores cases, jobs, results, audit events, and trial configuration records in separate collections with indexes selected for the main access paths: idempotent job creation, run and result lookup by trial, case, or slide, dashboard pagination, and startup restoration of active trials. The unique index on trial identifier and idempotency key is central to the execution model, since repeated scans over overlapping time windows should not resubmit the same slide/model input.

Adapters provide the boundary between the portable STEP workflow and institution-specific infrastructure. Data adapters return pathology cases and slides with stable identifiers, slide locations, status fields, and metadata; the package includes a \texttt{mock} adapter for development and a \texttt{sql\_lis\_example} adapter that demonstrates SQL-based LIS discovery patterns. Compute adapters encapsulate execution; the \texttt{local} adapter starts worker processes on the API host with configurable concurrency, whereas the \texttt{lsf} and \texttt{slurm} adapters submit one batch job per slide to HPC schedulers. Ancillary adapters support delayed linkage of external information to completed runs; the built-in \texttt{manual} adapter ingests validated JSON files by \texttt{run\_id}, while \texttt{none} disables this pathway. New adapters can be added by implementing the corresponding interface and registering a factory builder, leaving the orchestration logic unchanged.

The API layer (\texttt{step/api/main.py}) exposes the operational surface of the system. It supports trial registration and deregistration, manual triggering of scan and ingestion cycles, scheduler inspection, retry of failed runs, run-history queries, result views by case or slide, and read-only dashboard endpoints. This control plane allows STEP to be operated interactively during development while preserving the same execution path used by scheduled production-like trials.

\subsection{Process Life-cycle}

Figure~\ref{fig:pipeline}b summarizes the application life-cycle. A STEP trial begins with a \texttt{POST /silent-trials} request. The registration payload defines the trial identifier, scan cadence, scan window, data source, compute adapter, ancillary adapter, model name and version, ingestion intervals, and adapter-specific configuration. Unless explicit artifact directories are supplied, STEP creates trial-scoped payload, result, and ancillary directories under the configured artifact root. The resulting configuration is persisted and marked active, allowing the trial to be listed, stopped, or restored later.

Each scan uses the trial's configured lookback window to retrieve candidate cases and slides from the data adapter. The adapter is responsible for local discovery logic, but the returned objects follow a common internal representation. This lets the orchestrator treat a mock development source and a SQL-backed LIS source in the same way once cases have crossed the adapter boundary.

For every candidate slide, STEP computes an \texttt{input\_hash} from the slide identifier, slide location, model name, and model version, then combines it with the model version and slide identifier to form an idempotency key. If a matching job already exists for the trial, the slide is skipped for that scan. Otherwise, the orchestrator assigns a new \texttt{run\_id}, writes a payload JSON file, submits the job through the compute adapter, records the scheduler job identifier when available, and inserts the run as a queued job in MongoDB. This design makes repeated scans over overlapping accession windows safe while still allowing explicit retry of failed runs through a separate retry-specific idempotency key.

Inference execution is standardized by a worker contract rather than by a fixed model implementation. The runner invokes the configured entrypoint with a payload path and result directory, and the entrypoint writes one result JSON file named by \texttt{run\_id}. Required fields include the trial, run, case, and slide identifiers; model name and version; a score object; and a terminal or intermediate status. If the entrypoint fails or writes an invalid result, the runner can emit a standardized failed result so that ingestion still produces an auditable terminal state.

Result ingestion polls the selected compute adapter for completed artifacts and persists each result with its associated run. The repository updates job status, stores timing, error, score, metadata, and log references when present, and appends a result-ingestion audit event. Ancillary ingestion follows a parallel pattern: validated ancillary records are linked by \texttt{run\_id} and stored with the corresponding result when available. Together, the job, result, ancillary, and audit collections provide enough state to reconstruct what was discovered, what was submitted, what completed, what failed, and which external information was later attached.

\subsection{Platform Stack}

STEP is implemented in Python. Its current stack uses FastAPI and Uvicorn for the API service, APScheduler for recurring trial jobs, PyMongo for persistence, and pandas and pyodbc for tabular and SQL-oriented data access patterns. The package can be run against an external MongoDB service or with helper modes for local, Docker Compose, and HPC-oriented MongoDB startup workflows.

\section{Discussion}

STEP addresses an important gap in the translation of computational pathology AI: reusable infrastructure for evaluating models prospectively before they influence patient care. Unlike retrospective validation, silent trials require not only assessment of model performance but reliable case discovery, data access, inference execution, failure recovery, and preservation of operational provenance. STEP incorporates these functions into a common orchestration layer that can be reused across models and institutions.

A central design principle is the separation of trial orchestration from institution-specific infrastructure. STEP operates on common representations of cases, slides, runs, and results, while adapters encapsulate interactions with local laboratory information systems, storage environments, and compute backends. Across deployments, the same trial control flow, persistence model, and idempotency mechanisms can therefore be retained while site-specific integrations can be implemented at the adapter boundary. This suggests that a meaningful portion of the engineering required for prospective pathology AI evaluation can be standardized despite substantial heterogeneity in institutional infrastructure.

Persistent operational state is also important for evaluating deployment readiness. Failures in prospective systems can occur before, during, or after model inference: eligible cases may not be identified, slides may be unavailable, jobs may fail, or results may arrive outside clinically relevant time windows. STEP's run state and audit history support recovery from these failures while enabling measurement of operational endpoints such as case capture, inference success, failure rates, and turnaround time. In this respect, STEP complements rather than replaces model-serving or general-purpose workflow platforms; its scope is the trial-level orchestration and provenance required for prospective, non-interventional evaluation.

Several limitations remain. STEP does not establish clinical effectiveness, model safety, or generalizability, which require appropriate prospective analyses and ultimately interventional evaluation. Deployment also continues to require institution-specific development and validation of data integrations. Current compute adapters support local execution, LSF, and Slurm, while other schedulers, cloud environments, and container-orchestration platforms would require additional adapters. STEP also currently relies on internal adapter contracts rather than standards-based clinical interoperability. Finally, the deployments reported here primarily establish technical feasibility and portability; evaluation across additional institutions, models, and use cases will be needed to quantify implementation effort and assess broader generalizability.

Future development will focus on additional data and compute adapters, standards-based clinical data exchange, and richer monitoring and reporting of operational endpoints. By making silent-trial infrastructure reusable and auditable across heterogeneous pathology environments, STEP provides a practical foundation for evaluating computational pathology models under real-world conditions before interventional clinical deployment.

\section*{Declarations}

\textbf{Availability of Data and Materials:} Source code and documentation are available on \href{https://github.com/gabricampanella/STEP}{GitHub}.

\textbf{Competing Interests:} The authors declare that they have no competing interests.

\textbf{Authors' Contributions:} GC conceived and designed the STEP platform and drafted the manuscript. MC and OL implemented STEP at MSHS, with implementation support from JH and RK. PS supported the implementation at TUM, and CV supported the implementation at MSKCC. All authors reviewed and approved the final manuscript.

\textbf{Acknowledgements:} This work was supported in part through the Minerva computational and data resources and staff expertise provided by Scientific Computing and Data at the Icahn School of Medicine at Mount Sinai and supported by the Clinical and Translational Science Awards (CTSA) grant UL1TR004419 from the National Center for Advancing Translational Sciences.

\printbibliography

\end{document}